\documentclass{article}

\usepackage{arxiv}

\usepackage[utf8]{inputenc}

\usepackage[T1]{fontenc}

\usepackage[hidelinks]{hyperref}

\usepackage{url}

\usepackage{amsfonts}

\usepackage{graphicx}

\usepackage{multirow}

\usepackage{amsmath}

\usepackage{booktabs,array}

\usepackage{multirow}

\usepackage{stix,bbding,pifont,utfsym,fontawesome}

\usepackage{multicol}

\usepackage{makecell}

\usepackage{mathtools, cuted}

\usepackage{cite}

\usepackage{pgfplots}

\usepackage{subcaption}

\usepackage{balance}

\usepackage{tikz}

\usepackage{caption}

\usepackage{enumitem}
 
\pgfplotsset{compat = newest}

\usepackage{comment}

\usepackage[nameinlink]{cleveref}

\Crefname{figure}{Figure}{Figures}

\title{Towards understanding the performance of IEEE 802.11p MAC in heterogeneous traffic conditions}

\author{M S Gayathree \\
School of Computing and Electrical Engineering \\
Indian Institute of Technology Mandi \\
Himachal Pradesh, India \\
\texttt{s20005@students.iitmandi.ac.in} \\
\And
Sreelakshmi Manjunath \\
School of Computing and Electrical Engineering \\
Indian Institute of Technology Mandi \\
Himachal Pradesh, India \\
\texttt{sreelakshmi@iitmandi.ac.in} \\
}

\begin{document}

\maketitle

\begin{abstract}

Motivated by the need to study the performance of vehicular communication protocols as applicable to heterogeneous traffic conditions, we study the performance of IEEE 802.11p medium access protocol under such a traffic setup. We consider a setup comprising connected vehicles and human-driven Motorised Two Wheelers (MTWs), where the connected vehicles are required to move as platoon with a desired constant headway despite interruptions from the two wheelers. We invoke specific mobility models for the movement of the vehicles\textemdash car following models for connected vehicle platoons and gap-acceptance model to capture the movement of the MTWs\textemdash and use them to configure (i) the traffic setup and (ii) the rate at which data packets related to safety-critical messages need to be transmitted. A control-theoretic analysis of the car-following models yields a bound on the admissible communication delay to ensure non-oscillatory convergence of the platoon headway. We then use suitable Markov chain models to derive the distribution of the MAC access delay experienced by packets pertaining to safety-critical events as well as routine safety messages. The distribution along with the bound on the admissible delay enables us to derive the reliability of the 802.11p MAC protocol in terms of traffic and EDCA parameters. Our study highlights the need for redesign of MAC protocols for vehicular communications for safety-critical applications in heterogeneous conditions.

\end{abstract}


\section{Introduction}

Road safety remains a significant developmental challenge, a matter of public health concern, and a primary cause of fatalities and injuries globally. Annually, road traffic accidents result in the loss of about 1.3 million lives, with low and middle-income countries contributing to 93\% of the global road fatalities~\cite{statistics2}.
Connected vehicles and Intelligent Transportation Systems (ITS) are expected to play a vital role in the future of road transportation. Equipped with vehicular communication, connected vehicles are capable of planning and performing cooperative movement for improved safety, better ride quality and shorter ride times. This ability to perform cooperative movement is dependent on the efficiency of vehicular communication protocols. Therefore, understanding the performance of these protocols is central to the design of connected vehicles and ITS. 
Further, when deployed, particularly in developing nations, connected vehicles would have to co-exist and interoperate with conventional vehicles of various types and dimensions, and with varying levels of automation. This necessitates the study of vehicular communication protocols in the context of heterogeneous traffic conditions. 

Platooning, arguably the most basic cooperative movement of vehicles, refers to vehicles moving in a single lane to maintain a constant spacing between each other. 
Such cooperative movement requires decentralised decision making and is hence sensitive to time delays. Connected vehicles deploy the IEEE 1609 protocol stack, and communicate using Dedicated Short Range Communication (DSRC). At the MAC layer, the IEEE 802.11p standard with Enhanced Distributed Channel Access (EDCA) mechanism is used. In congested environments, with large number of vehicles, large values of channel access delays can cause huge packet delays, thus impacting time-critical applications such as platooning. Hence, it is desirable to study the performance of this EDCA mechanism in terms of delay.

Performance analysis of the IEEE 802.11p MAC protocol has indeed attracted significant research attention. 
In~\cite{ieee1}, the performance of IEEE~$802.11$p EDCA is analyzed using a~$2$-D Markov chain model that considers varying Contention Window~(CW), internal collision scenarios and freezing mechanism. The performance metrics, including normalized throughput and time delay, has been calculated and derived. An analytical model for the IEEE~$802.11$p EDCA mechanism, has been evaluated through extensive simulations and introduced in~\cite{ieee2}. The model was created by merging two Markov chain models \textemdash~a $2$-D model representing the backoff procedure of each Access Category~(AC) queue and a $1$-D model describing contention period resulting from varying Arbitrary Inter-Frame Space~(AIFSs) and Contention Windows~(CWs).

In~\cite{highway}, a comprehensive analytical model for assessing the performance of the protocol has been developed. The model incorporates two Markov chains for different priority ACs, which are later used to analyze the delay distribution in broadcast mode. The accuracy of the model in its prediction of standard deviation, mean, and probability distribution of the delay is verified through simulations conducted on Network Simulator~$2$~(NS$2$). Further, the idea was extended to compute the reliability of IEEE~$802.11$p in terms of packet delay and packet delivery ratio in~\cite{highway_reliability}. In the other hand, in~\cite{OnMACAccess}, a dual-slope propagation model is employed to determine the critical ranges in vehicular communications such as the carrier sensing range and interference range, which are typically assumed to be constant. Further, the authors of~\cite{highway,OnMACAccess},  established that the shifted exponential distribution can serve as an effective approximation of the MAC access delay. A performance analysis of IEEE~$802.11$p for continuous backoff freezing is conducted in~\cite{backofffreezing}. In~\cite{hidden}, IEEE~$802.11$p MAC sublayer is studied in the presence of hidden terminals. The backoff procedure for each AC is modeled using a~$3$-D Markov chain. Additionally, the model considers data packets and control packets separately.
Most of these studies adopt the approach of mathematical modelling and analysis, followed by simulations, for performance analysis. 
A comparison of the existing work is provided in Table \ref{tab:Table 1}.
Of these studies, only~\cite{23,25,1} consider mobility models for vehicular movement, while the others assume that the vehicle positions follow certain stochastic processes. 
Notably, all these studies focus on homogeneous traffic settings with a common assumption that all the vehicles are equipped with vehicular communication. Performance of the IEEE 802.11p in the context of heterogeneous traffic environments, and its reliability for specific cooperative movement such as platooning in such scenarios merits investigation. 

\begin{figure*}[!h]
\centering
\scalebox{0.8}{
\begin{tikzpicture}

\draw (0,0) rectangle (11,4.5);
\node at (5.5,4.75) {\textbf{Traffic Setup}};

\node at (2.25,3.95) {\underline{Mobility Model}};
\node[rectangle, draw, text width=4cm] at (2.25,2.25) {\begin{center} \vspace{-0.25cm} Platoon of cars \end{center}
\begin{itemize}
\item FVD, MOVM Model
\item Packets generated at a constant rate $\lambda_1$ pkts/sec
\end{itemize}};

\node at (8.75,4.1) {\underline{Heterogeneous Traffic}};
\node[rectangle, draw, text width=4cm] at (8.75,2.25) {2-wheeler interrupting the platoon of cars
\begin{itemize}
\item Modelled using Gap Acceptance Model
\item Packets generated at rate $\lambda_0$ pkts/sec
\end{itemize}};

\draw[->, thick, black] (4.375,2.25) -- (6.625,2.25);
\node at (5.5,3) {Eqm. headway};
\node at (5.5,2.5) {($y^\ast$)};

\draw[->, thick, black] (11,2.25) -- (14.37,2.25);
\node at (12.695,3) {Rate of pkt generation};
\node at (12.695,2.5) {($\lambda_0$, $\lambda_1$)};

\node at (17,3.9) {\textbf{V2V Communication}}; 
\node[rectangle, draw, text width=5cm] at (17,2.25) {\begin{center} \vspace{-0.25cm} \underline{IEEE 802.11p Protocol}\\ \vspace{0.25cm}Obtain mean MAC access\\delay by performing: \end{center}
\begin{enumerate}[label=(\roman*)]
\item Mathematical Analysis
\item Protocol Simulation  
\end{enumerate}};

\draw[->, thick, black] (2.25,0.825) -- (2.25,-1.175);
\draw[->, thick, black] (17,0.875) -- (17,-1.375);

\node[rectangle, draw, text width=4cm] at (2.25,-2) {\begin{center} \vspace{-0.25cm} \underline{Local Stability Analysis} \end{center} \begin{center} Bound on delay for non-oscillatory convergence \end{center}};

\node[rectangle, draw, text width=7cm] at (17,-2) {\begin{center} \vspace{-0.25cm} \underline{Reliability} \end{center} 
\begin{center} \(\mathbb{P}\) (MAC Access Delay < Critical Delay) \end{center}};

\draw[->, thick, black] (4.375,-2) -- (13.375,-2);
\node at (9.000,-1.75) {Critical Delay};
\end{tikzpicture}}

\captionof{figure}{Overview of our work}
\label{overview}
\end{figure*}
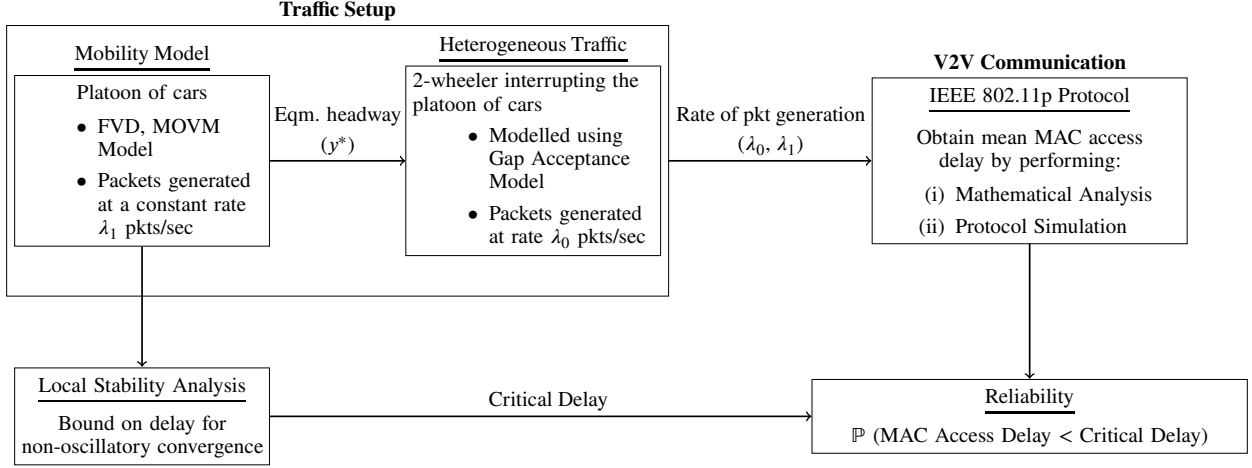

This paper presents a study aimed at understanding the performance of the IEEE 802.11p medium access protocol, for platooning of connected vehicles in the presence of motorised two wheelers. We consider the (i) Full Velocity Difference (FVD) model and the (ii) Modified Optimal Velocity Model (MOVM)~\cite{9} for the connected vehicle platoon and the gap acceptance model~\cite{gap} to capture the probability that a two wheeler interrupts the platoon by taking up the gap between two consecutive connected vehicles. These models are used to configure the traffic setup and the rate at which safety-critical/routine messages need to be sent, \emph{i.e.} the rate at which data packets need to be transmitted. Stability analysis of the FVD and MOVM yields a bound on the time delay that serves as the necessary and sufficient condition for non-oscillatory convergence of the platoon headways , \emph{i.e.} for each vehicle the headway converges to a desired constant without exhibiting any oscillations. We define the reliability of IEEE 802.11p as the probability that the delay experienced by a given packet is within the critical delay defined by the aforementioned bound. We use MATLAB to carry out the mathematical analysis and protocol simulation to evaluate the mean packet delay and reliability, for various headway requirements of the platoon. We consider two specific cases for generation of packets corresponding to safety-critical events: packets are generated as per a Poisson process with (i) a constant rate, (ii) rate as a function of the platoon headway, based on the probability of a two wheeler interrupting the platoon thereby generating a safety-critical event. Figure~\ref{overview} gives an overview of our work.

The rest of the paper is organised as follows. The mobility models are described in Section~\ref{sec:Mobility_models}. The performance evaluation of 802.11p using the mathematical model and simulation results are outlined in Section~\ref{sec:performance evaluation}. The conclusions and future work are discussed in Section~\ref{sec:Contributions}.

\begin{table}[tbh]
\caption{Comparison of existing studies}
\centering
\begin{tabular}{ccccc}
\toprule
References & Steady/Time Varying & Platoon & Mobility Model & Traffic Type \\
\toprule
\cite{highway_reliability} & \makecell{Steady-state} & \makecell{$\times$} & \makecell{Not considered} & \makecell{Homogeneous} \\
\midrule
\cite{OnMACAccess} & \makecell{Steady-state} & \makecell{$\times$} & \makecell{Not considered} & \makecell{Homogeneous} \\
\midrule
\cite{23} & \makecell{Steady-state} & \makecell{$\checkmark$} & \makecell{Intelligent Driver Model} & \makecell{Homogeneous} \\
\midrule
\cite{16} & \makecell{Steady-state} & \makecell{$\times$} & \makecell{Constant Position Model of NS3} & \makecell{Homogeneous} \\
\midrule
\cite{20} & \makecell{Steady-state} & \makecell{$\times$} & \makecell{Not considered} & \makecell{Homogeneous} \\
\midrule
\cite{25} & \makecell{Time-varying} & \makecell{$\times$} & \makecell{Freeway Mobility Model} & \makecell{Homogeneous} \\
\midrule
\cite{1} & \makecell{Time-varying} & \makecell{$\checkmark$} & \makecell{Intelligent Driver Model} & \makecell{Homogeneous} \\
\midrule
\cite{27} & \makecell{Time-varying} & \makecell{$\times$} & \makecell{Not considered} & \makecell{Homogeneous} \\
\midrule
\makecell{Our Work} & \makecell{Steady-state} & \makecell{$\checkmark$} & \makecell{FVD, MOVM \& Gap Acceptance Model} & \makecell{Heterogeneous} \\
\bottomrule
\end{tabular}
\label{tab:Table 1}
\end{table}

\section{Traffic setup and Mathematical models}
\label{sec:Mobility_models}

We now describe the traffic setup considered in our work, and briefly discuss the priority-based packet classification used in the IEEE 802.11p MAC protocol. 
This is followed by a description of the mobility models for the connected vehicle platoon and the probability that a two wheeler maneuvers into the platoon to fill the gap caused by the headway between two consecutive vehicles in the platoon. We then use this \emph{gap-acceptance} probability  to model the rate of generation of packets corresponding to event-driven traffic.

\subsection{Traffic setup and vehicular communication}
\label{sec:TrafficSetup}

The objective of our work is to study the performance of the IEEE 802.11p protocol in the context of platooning of connected vehicles in heterogeneous traffic conditions. To that end, we consider a traffic setup that comprises two types of vehicles; namely connected vehicles and human-driven motorised two wheelers. The connected vehicles are assumed to be traversing on a single-lane road, as a platoon, with no overtaking. Human-driven motorised two wheelers are assumed to surround the platoon throughout the road, and may interrupt the platoon by overtaking, filling the gap or cutting across the platoon. The connected vehicles must maintain their platoon despite these interruptions by the two wheelers. 

The IEEE 802.11p protocol allows prioritised channel access in order to ensure that packets pertaining to safety-related applications are prioritised over others. Generally, there are four access categories; namely AC0 -- AC3, with AC0 being the highest priority category and AC3 being the lowest. Of these, AC0 and AC1 are reserved for safety-related applications while the other two may be used for infotainment. Among the two, it is a general practice to reserve the AC0 category for safety-critical messages and allot AC1 category for routine safety-related applications.
 
\begin{figure}
\centering
\begin{tikzpicture}
\scalebox{0.9}{
\draw [line width = 1mm, line cap=round](-1,0) -- (17,0);
\draw [line width = 1mm, line cap=round](-1,-2.5) -- (17,-2.5);

\node[inner sep=0pt] at (0,-1) {\includegraphics{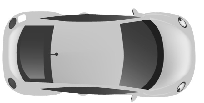}};
\node[inner sep=0pt] at (4,-1) {\includegraphics{2.eps}};
\node[inner sep=0pt] at (8,-1) {\includegraphics{2.eps}};
\node[inner sep=0pt] at (12,-1) {\includegraphics{2.eps}};
\node[inner sep=0pt] at (16,-1) {\includegraphics{2.eps}};
\node[inner sep=0pt] at (6,-1) {\includegraphics{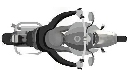}};

\node[inner sep=0pt] at (2,-2) {\includegraphics{1.eps}};

\draw [line width = 0.5mm,dashed,->] (2.6,-2) .. controls (4.5,-2) .. (5.75,-1.25);

\draw [line width = 0.75mm,<->] (12.975,-1) -- (15.025,-1);
\node[text width=0.5cm,fill=white] at (14,-1) {\huge $y^{\ast}$};
}
\end{tikzpicture}
\caption{Schematic diagram of the heterogeneous traffic setup under study.}
\label{fig:my_label}
\end{figure}

In order to ensure safe platooning, the connected vehicles must communicate their position/velocity information to other connected vehicles in their vicinity. Therefore, each vehicle needs to send packets periodically at a pre-defined rate. Additionally, in the heterogeneous traffic setting that we consider, each vehicle will have to communicate its position/velocity as and when a two wheeler interrupts with the platoon. The rate at which these packets are generated would have to depend on the probability of such an interruption. As such interruptions can be considered safety-critical, we classify packets pertaining to these events as high priority AC0 category, while the periodic updates are assigned the lower priority AC1. Figure~\ref{fig:my_label} illustrates the traffic setup. Here, we assume that the connected vehicles encounter no other safety-critical events, apart from the interruptions by the motorised two wheelers. Therefore, packets pertaining to the AC1 traffic are generated periodically at pre-defined rate; and those of the AC0 traffic are event-based and are generated based on an appropriate mathematical model. 

\subsection{Connected vehicle platoon: Car following models}
\label{sec:Platoon_MOVM}

The Optimal Velocity Model (OVM)~\cite{6} is one of the earliest models for platooning, in which each vehicle updates its velocity according to the position of the vehicle ahead. The position of the preceding vehicle can either be obtained through driver perception or through V2V communication. In case of the former, the model must account for the reaction delay of the driver, while for the latter communication delays must be accounted for. In addition to these delays, one must also consider mechanical delays such as actuator delay, braking delay etc. The OVM models the acceleration of a vehicle as a function of its headway, \emph{i.e} the distance between the preceding vehicle and itself.
In this work, we consider two variants of the OVM, namely the (i) Modified Optimal Velocity Model (MOVM)~\cite{8} and the (ii) Full Velocity Difference (FVD) model. 
The MOVM, like the OVM, describes the acceleration of a vehicle as a function of its headway. However, unlike the OVM it accounts for the reaction delay of the driver, which makes the feedback regarding the vehicle's headway delayed. The FVD, on the other hand, considers the influence of the velocity difference between itself and the preceding vehicle in addition to the influence of its headway. This model too can be enhanced to consider the delayed feedback. Which is the approach we take in our work. 

For a platoon of $N$ traversing on an infinite highway, with the relative velocities and headways chosen as the state variables, the FVD can be written as 
\begin{equation}
\begin{aligned}
    \dot{v}_1 (t) =&\, \ddot{x}_0 (t) + a (\dot{x}_0 (t) - V (y_1 (t - \tau) ) - v_1 (t-\tau) ) - lv_1(t-\tau), \\
    \dot{v}_k (t) =&\, a (V (y_{k-1} (t - \tau) ) - V (y_k (t - \tau) ) - v_k (t-\tau) ) +\, l\left(v_{k-1}(t-\tau-v_k(t-\tau))\right), \\
    \dot{y}_i (t) =&\, {v}_i (t)
\end{aligned} \label{eq:FVD}
\end{equation}
for ~$i = 1, 2, \dots , N$ and for~$k = 2, 3, \dots , N$. Here, $a$ and $l$ are the  sensitivity parameters, $x_0 (t), \dot{x}_0(t)$ and $\ddot{x}_0(t)$ represent the position, velocity and acceleration, respectively, of the lead vehicle. Further, $v_i(t)$ and $y_i(t)$ indicates the relative velocity and the headway between the $i^{th}$ and ${(i-1)} ^{th}$ vehicle, respectively. The feedback delay $\tau$ represents the cumulative delay in the communication channel and other delays such as actuation and braking delays in the vehicle. Note that for the sake of this analysis, we make the simplifying assumption that the delay experienced by all the vehicles is the same, $\tau$. The function $V(\cdot)$ represents the Optimal Velocity Function (OVF), which models the acceleration of a follower vehicle as a function of the available headway. One such OVF, known as the Bando OVF model is given by~\cite{6}  
\begin{equation}
  V(y) = V_0 \left( \tanh \left( \frac{y - y_m}{\tilde{y}} \right) + \tanh \left( \frac{y_m}{\tilde{y}} \right) \right)\label{18}
\end{equation}
where $V_0$, $y_m$ and $\tilde{y}$ are model parameters. The model~\eqref{eq:FVD} boils down to the MOVM for the case $l=0$. While the FVD is known to be a more accurate representation for a platoon of autonomous vehicles, the MOVM captures the scenario of human-driven vehicles better. 
The analysis that follows would apply in either case subject to appropriate choice of the parameter $l$. 

Recall that for successful platooning, we require that the spacing between each consecutive pair of vehicles, \emph{i.e.} each vehicle's headway, be constant. This condition also implies that the relative velocities of the vehicles must converge to zero. This would ensure collision-free platooning and jerk-less movement of each vehicle.  For a platoon represented by system~\eqref{eq:FVD}, successful platooning can be guaranteed when the headways $y_i(t)$ converge to a constant value and the relative velocities $v_i(t)$ converge to zero. In systems-theoretic terms, we require the dynamical system~\eqref{eq:FVD} to converge to a stable equilibrium. Generally, in order to find the conditions on the system parameters that ensure convergence to stable equilibrium one would carry out a local stability analysis. Such analysis and the resulting analytical conditions can be found in~\cite{yu2013full} for the FVD model and in~\cite{8} for the MOVM. 

While such conditions for local stability indeed ensure that the headways converge to a constant, in practical situations it is often desired that the headway does not undergo any oscillations while it converges. Such \emph{non-oscillatory} convergence ensures smooth and jerk free movement of the vehicles in the platoon. We now derive a bound on the time delay $\tau$ that ensures non-oscillatory convergence of the platoon headways. Below we present the analysis for the FVD model, from which one can also derive required bound for the MOVM by substituting $l=0.$
The ensuing analysis closely follows~\cite{9}. 

Equating the first derivative of the state variables in system~\eqref{eq:FVD} to zero yields the equilibrium point $y_i^{\ast}$ = $V^{-1}(\dot{x})$ and $v_i^{\ast}=0.$ Given the velocity of the lead vehicle at equilibrium, i.e., $\dot{x}_0$, the equilibrium headway, $y_i^{\ast}$ can be obtained for given OVF parameters. Consider a perturbation from the equilibrium headway $u_{i} (t)$ = $y_{i} (t)$ - $y_i^{\ast} (t)$. For small values of the perturbation, we may take a Taylor series expansion about the equilibrium to obtain a linearised approximation of~\eqref{eq:FVD}. This yields
\begin{equation}
\begin{aligned}
    \dot{v}_1 (t) =&\, -aV^{\prime}(y^\ast) u_1 (t - \tau) - (a+l) v_1 (t-\tau), \\
    \dot{v}_k (t) =&\, aV^{\prime}(y^\ast) \left(u_{k-1} (t - \tau) - u_k (t - \tau)\right) - a v_k (t-\tau) +\, l(v_{k-1}(t-\tau)-v_k(t-\tau)), \\
    \dot{u}_i (t) =&\, v_i (t),
\end{aligned} \label{taylorseries}
\end{equation}
where prime denotes differentiation with respect to the state variable.
The above linearised system has the following characteristic equation 
\begin{equation}
    f (\lambda) =  \big(\lambda^2 +  (a + l) \lambda e^{-\lambda \tau} + aV'(y^\ast) e^{-\lambda \tau}\big)^N = 0, \label{lambda}
\end{equation}
which can simply be reduced to 
\begin{equation}
    f (\lambda) =  \lambda^2 +  (a + l) \lambda e^{-\lambda \tau} + aV'(y^\ast) e^{-\lambda \tau} = 0, \label{eq:char_eq}
\end{equation}
In order to obtain the necessary and sufficient condition for non-oscillatory convergence, we look for the bound on the delay $\tau$ and the system parameters that ensures that the roots of the characteristic equation~\eqref{eq:char_eq} lies on the negative real axis. To find this, we begin by assuming $\lambda = \sigma+i\omega,$ and then deriving the condition that ensures $\omega =0.$ 

Substituting $\lambda = \sigma+i\omega$ in~\eqref{eq:char_eq} and separating the real and imaginary parts, we obtain
\begin{align*}
    \left(\sigma(a+l) + aV'(y^\ast)\right)e^{-\sigma\tau}\,\cos(\omega\tau) + (a+l)\omega e^{-\sigma\tau}\sin(\omega\tau) =&\, \omega^2-\sigma^2,\\
    (a+l)\omega e^{-\sigma\tau}\cos(\omega\tau) - \big(\sigma(a+l) + aV'(y^\ast)\big)e^{-\sigma\tau}\,\sin(\omega\tau) =&\, -2\sigma\omega.
\end{align*}
Squaring and adding these equations, we obtain
\begin{align}
\left(\sigma(a+l) + aV'(y^\ast)\right)^2 + \big((a+l)\omega\big)^2 =\, \big((\omega^2-\sigma^2)^2+4\sigma^2\omega^2\big)e^{2\sigma\tau}\label{eq:sqaresum_of_char}
\end{align}
For $\lambda$ to be a real root, we require $\omega=0$. Substituting $\omega=0$ in~\eqref{eq:sqaresum_of_char} yields
\begin{align*}
\left(\sigma(a+l) + aV'(y^\ast)\right)^2 = \sigma^4 \, e^{2\sigma\tau},
\end{align*}
which serves as the necessary condition for non-oscillatory convergence. For this to also be the sufficient condition, we must ensure that $\omega=0$ serves as unique solution to~\eqref{eq:sqaresum_of_char}.  Substituting for $\left(\sigma(a+l) + aV'(y^\ast)\right)^2$ in equation~\eqref{eq:sqaresum_of_char}, we get
\begin{align*}
\sigma^4e^{2\sigma\tau} + (a+l)^2\omega^2 = (\omega^2+\sigma^2)^2e^{2\sigma\tau},
\end{align*}
which upon simplification yields
\begin{align*}
\omega^2\big(\omega^2+2\sigma^2-(a+l)^2e^{-2\sigma\tau}\big)=0.
\end{align*}
For $\omega=0$ to be a unique solution, we require $2\sigma^2\,e^{2\sigma\tau}=(a+l)^2$. Therefore, the characteristic equation~\eqref{eq:char_eq} has purely real roots if and only if 
\begin{align*}
\left(\sigma(a+l) + aV'(y^\ast)\right)^2 = \sigma^4 \, e^{2\sigma\tau} && \text{and} && 2\sigma^2\,e^{2\sigma\tau}=(a+l)^2.
\end{align*}
Using these two conditions and solving for $\sigma$ yields
\begin{align*}
\sigma = \tilde{d}(-2\pm\sqrt{2}),
\end{align*}
where $\tilde{d} = aV'(y^\ast)/(a+l)$. 
Substituting this value of $\sigma$ and $\omega=0$ in the characteristic equation~\eqref{eq:char_eq} yields the critical value of $\tau$ which marks the necessary and sufficient condition for non-oscillatory convergence. We denote this critical value by $\tau_{cr}$, and this value is given by 
\begin{align}
\tau_{cr} = \frac{1}{\tilde{d}(-2-\sqrt{2})}\ln\bigg(\frac{-\tilde{d}(a+l)(-2-\sqrt{2})-aV'(y^\ast)}{\big(\tilde{d}(-2-\sqrt{2})\big)^2}\bigg).\label{eq:tau_noc}
\end{align}
Therefore, any value of feedback delay that satisfies $\tau <=\tau_{cr}$ ensures non-oscillatory convergence. 
%
%
%
%
Note that the value of critical delay depends on the model parameters $a, V_0, y_m, \tilde{y}$.  
 
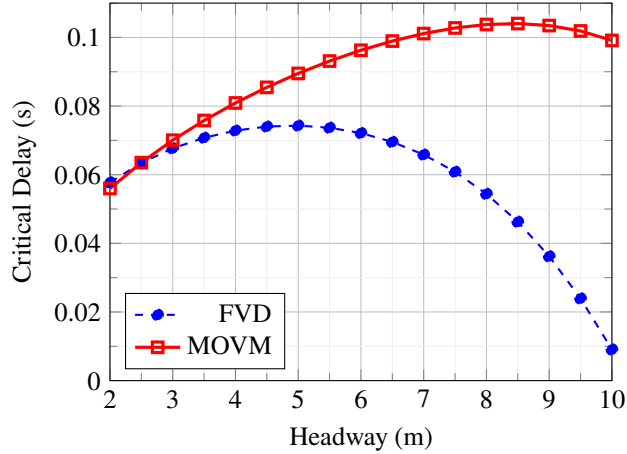
\begin{figure}[!h]
\centering 
\begin{tikzpicture}
\begin{axis}[
    yticklabel style={
      /pgf/number format/fixed,
      /pgf/number format/precision=2,
      /pgf/number format/fixed zerofill,
      /pgf/number format/fixed relative,
    },
    every axis plot/.append style={thick},
    xlabel = {Headway (m)},
    ylabel = {Critical Delay (s)},
    xmin = 2, xmax = 10,
    ymin = 0, ymax = 0.11,
    xtick distance = 1,
    ytick distance = 0.02,
    grid = both,
    minor tick num = 1,
    major grid style = {lightgray},
    minor grid style = {lightgray!25},
    width = 0.5\textwidth,
    height = 0.4\textwidth,
    legend cell align = {right},
    legend pos = south west
]
 
\addplot[dashed,blue,mark=*,style=thick]  table [x = {x}, y = {y}] {critical_delay_FVD.dat};
\addplot[red,mark=square,style=very thick]  table [x = {x}, y = {y}] {critical_delay_OVM.dat};
 \legend{
    FVD , 
    MOVM
    }
 \end{axis}
\end{tikzpicture}
\caption{Critical delay for non-oscillatory convergence of the platoon headways as a function of the equilibrium headway ($y^\ast$) computed using equation~\eqref{eq:tau_noc}. } 
\label{fig:CriticalDelay_Headway}
\end{figure}

We now present a graphical representation of the critical delay for non-oscillatory convergence for the two models that we consider. For our study, we consider $\tilde{y}= 10$, $y_m = 5,$ as in~\cite{9} and $a = 5$  while $\dot{x}_0$ is set to be $25$ m/s. For FVD, we consider $l=2,$ and for the MOVM we use $l=0$. The desired headway is varied in the range $[2,10]$ m, and the corresponding value of $V_0$ that satisfies the equilibrium condition $y_i^\ast = V^{-1}(\dot{x_0})$ is calculated. Using these parameter values, we compute the value of the critical delay using the condition~\eqref{eq:tau_noc}. The plot showing the critical delay as a function of the headway is presented in Figure~\ref{fig:CriticalDelay_Headway}. The region below the curve represents the values of time delay for non-oscillatory convergence of the platoon. 

So far we have derived a bound on the feedback delay that ensures non-oscillatory convergence of the platoon headway. The feedback delay, in general, is an accumulation of the communication delay, the delay in the on-board computers and the throttle, brake and steering actuator delays. Studies suggest that the computational and mechanical delays are  much larger than communication delays, and typically comprise 85-90\% of the total feedback delay~\cite{beregi2021connectivity}. To that end, we consider the bound on packet delay as  $10\%$ of the critical delay for non-oscillatory convergence.

\subsection{Two Wheeler Maneuver: Gap Acceptance Model}
\label{sec:TwoWheeler_GapAcceptance}

In a heterogeneous traffic setting, one must consider the possibility of other vehicles interrupting the platoon\textemdash by overtaking, cutting across the platoon or taking up the gaps in the platoon. Such interruptions can be viewed as safety-critical events that are required to be communicated by vehicle experiencing it to the other vehicles in the platoon. The frequency of these safety-critical events would thus govern the number of packets corresponding to the event-driven (AC0) traffic that need to transmitted by the connected vehicle. We now outline a model that captures the event of a motorised two wheeler interrupting with the connected vehicle platoon. Further, we use it to model the rate at which event-driven traffic is generated at each transmitting vehicle.

For our work, we consider a gap acceptance model that is shown to be applicable specifically to the context of heterogeneous traffic in India. Probability that a two wheeler will accept gap between any two consecutive cars in the platoon is given by~\cite{gap}
\begin{equation}
  P_i = \frac{\exp{\left(\alpha + \beta_0\, \frac{y_i^{\ast}}{\dot{x}_0} \right)}}{1+\exp{\left(\alpha + \beta_0\, \frac{y_i^{\ast}}{\dot{x}_0} \right)}}\label{gap}
\end{equation}
where, $y_i^{\ast}$  is the equilibrium headway of the platoon and $\dot{x}_0$ is the velocity of the vehicles in the platoon. The ratio $y_i^{\ast}/\dot{x}_0$, often termed as time gap, is the time that a vehicle in the platoon will take to cover the headway distance. Further, $\alpha$ and $\beta$ are the model parameters that must be calibrated according to the type of the vehicle that is considered to be filling the gap in the platoon. For the case of motorised two wheelers interrupting with a platoon of cars, 
it is proposed that $\alpha = -1.933$ and $\beta_0 = 0.652$~\cite{gap}.

Equipped with the probability of occurrence of a safety-critical event, we now model the rate of packet generation for the AC0 category. To begin with, as these packets are generated by events, we assume that the packets are generated as per a Poisson process. Let the rate of this process be $\lambda_0$ pkts/sec. 
It is intuitive that $\lambda_0$ would depend on the gap-acceptance probability given by~\eqref{gap}. Further, with increase in the probability of two wheelers interrupting the platoon, the rate $\lambda_0$ must increase. Hence, $\lambda_0$ can be modelled as a non-decreasing function of $P_i$. The non-decreasing functions that have been used to model AC0 traffic in our study are listed in Table \ref{tab:Table 2}. Here, $k$ is a tunable parameter. We use $k=500$. 

\begin{table}[!h]
\centering
\caption{Mathematical models for rate of generation of AC0 packets ($\lambda_0$) as a function of gap acceptance probability ($P_i$)}
\label{tab:Table 2}
\begin{tabular}{ccccc}
\toprule
Non--decreasing Function & $\lambda_0(P_i)$ \\
\toprule
\toprule
Linear & $k P_i$ \\\\
Quadratic & $k \left( P_i^2 + P_i \right)$ \\\\
Sigmoidal & $k \left ( \frac{e^{P_i} - e^{-P_i}}{e^{P_i} + e^{-P_i}} \right)$ \\\\
Logarithmic & $1 - k\,\log \left( 1 - P_i \right)$ \\
\bottomrule
\end{tabular}
\end{table} 

\section{Performance Evaluation}
\label{sec:performance evaluation}

Consider a straight single-lane road with connected vehicles spaced as per the desired headway, which need to communicate their position/velocity information in order to maintain the platoon. We consider a single platoon as in Figure~\ref{fig:my_label}. These vehicles try to access the communication channel using the IEEE 802.11p protocol, and send the packets in broadcast mode wherein a transmitted packet is expected to be received by all vehicles within a pre-defined range. In a single vehicle, both AC0 and AC1 compete to access the communication channel. However, only one of them can access the channel at any given time. Each AC maintains its own Contention Window (CW), and if both ACs try to access the channel simultaneously, a virtual collision occurs. In this scenario, AC0 takes priority and is transmitted, while AC1 enters a backoff stage with its CW doubled. This process continues until AC1 gets a chance to access the channel, or until it reaches the maximum retransmission limit, at which point the packets will be discarded. As a result, only one vehicle (with only one AC) can access the channel within a given time slot to transmit information to its neighbouring vehicles.

Let $M$ be the maximum number of times that the CW of $AC_1$ is doubled. Then, $M$ can be given as
\begin{equation}
    M = \log_2 \frac{CW_{1,max} + 1}{CW_{1,min} + 1} \label{1}
\end{equation}
Then, the CW of $AC_1$ in the $j^{th}$ backoff stage is given by:
\begin{align}
    W_{1,j} = \begin{cases}
    2^j W_{1,0} ~~, j\leq M \\
    2^{M} W_{1,0} ~~,  < j \leq L
\end{cases} \label{2}
\end{align}
where, $W_{1,j}$ is the maximum CW size of $AC_1$ in the $j^{th}$ backoff stage and $L$ is the retry limit of $AC_1$.
For each AC within a vehicle, if the channel is idle for an amount of time equal to an Arbitrary Inter-Frame Space (AIFS), it transmits. In any other case, the AC will continue to monitor the channel until it is idle for a period of time up to AIFS. The AIFS for the two AC's is given as:
\begin{equation}
    AIFS[i] = SIFS + AIFSN[i] \times \sigma ~~, i=0, 1 \label{3}
\end{equation}
where, $SIFS$ is the Short Inter-Frame Space and $\sigma$ represents one slot time.\\
Since AC1, being of lower priority will have to wait more compared to AC0 to start contending for the channel, the Arbitration Inter-Frame Space Number (AIFSN) for AC1 is greater than that of AC0, i.e., $AIFSN1 > AIFSN0$. The difference between the two is given by $A_1$ \emph{i.e.}
\begin{equation}
    A_1 = AIFSN[1] - AIFSN[0] \label{ai equation}
\end{equation}
Let $\omega_i$ denote the internal transmission probability and $p_{vi}$ represent the virtual collision probability of $AC_i$. Then,
\begin{equation}
\begin{cases}
    p_{v0} = 0 \\
    p_{v1} = \omega_0
\end{cases} \label{4}
\end{equation}
The external transmission probability i.e $\tau_i$ is given as:
\begin{equation}
\begin{cases}
    \tau_0 = \omega_0 \\
    \tau_1 = \omega_1 (1 - p_{v1}) = \omega_1 (1 - \omega_0)
\end{cases} \label{5}
\end{equation}

Then, the total transmission probability is the sum of the internal transmission probabilities given by:
\begin{equation}
    \tau = \tau_0 + \tau_1 \label{6}
\end{equation}
The average transmission time of a packet is given by:
\begin{equation}
    T_{tr} = \frac{PHY_H}{R_b} + \frac{MAC_H + E[P]}{R_d} + \delta \label{7}
\end{equation}
where, $PHY_H$ and $MAC_H$ are the header length of the physical and MAC layer respectively. $\delta$ is the propagation delay, $R_b$ is the basic rate, $R_d$ is the data rate and $E[P]$ is the mean packet size. 
The AC0 messages arrive with a Poisson point process (with rate $\lambda_0$) whereas the AC1 messages arrive periodically at a rate of $\lambda_1$ (packets per second). 
The packet arrival probability of AC0 and AC1 is given as:
\begin{equation}
\begin{cases}
    p_{a0} = \sum \limits_{k = 1}^{\infty} \frac{(\lambda_0 \sigma)^k}{k!} e^{-\lambda_0 \sigma} = 1 - e^{-\lambda_0 \sigma}\\
    p_{a1} = \lambda_1 \sigma 
\end{cases} \label{8}
\end{equation}
Higher priority AC0 can be modelled as a one-dimensional Markov chain and the lower priority AC1 can be modelled as a two-dimensional Markov chain~\cite{highway_reliability}.\\
$N_{cs}$ is the number of vehicles which are in the carrier sensing range of the given vehicle. $p_{bi}$ is the backoff blocking probability, \emph{i.e.}, the probability that the channel is occupied by the other surrounding vehicles or there is a transmission attempted by the other ACs within the same vehicle. $p_{bi}$ is given by:
\begin{equation}
\begin{aligned}
  p_{bi} = 1 - \left[ P(k) \prod_{\substack{j = 0 \\ j \neq i}}^{1} (1 - \omega_j) \right]^{A_i + 1} ~~, i=0, 1 \\
  P(k) = \sum_{k=0}^{\infty} (1 - \tau)^k \frac{(N_{cs} - 1)^k}{k!} e^{-(N_{cs} - 1)}
\end{aligned} \label{9}
\end{equation}

$\rho_i$ is the server utilization factor of $AC_i$ and is given as:
\begin{equation}
    \rho_i = \frac{\lambda_i}{\mu_i} ~~, i = 0, 1 \label{11}
\end{equation}

Using the one-step transition probabilities, it is possible to express all stationary probabilities as $b_{0,0}$ and $b_{1,0,0}$~\cite{highway_reliability} \emph{i.e.},
\begin{equation}
\begin{cases}
b_{k} = \frac{(W_{0,0}-k)}{W_{0,0}(1-p_{b0})}b_{0} \\
b_{IDLE0} = \frac{1-\rho_{0}}{p_{a0}}b_{0}\\
\end{cases}
\end{equation}

\begin{equation}
\begin{cases}
b_{1,j,0} = p_{v1}^{j}b_{1,0,0}, ~~~j \in (1,L) \\
b_{1,0,k} = \frac{(W_{1,0}-k)}{W_{1,0}(1-p_{b1})}b_{1,0,0}, ~~~k \in (1,W_{1,0}-1)\\
b_{1,j,k} = \frac{(W_{1,j}-k)p_{v1}^{j}}{W_{1,j}(1-p_{b1})}b_{1,0,0}, ~~~j \in (1,M), ~~~k \in (1,W_{1,j}-1)\\
b_{1,j,k} = \frac{(W_{1,M}-k)p_{v1}^{j}}{W_{1,M}(1-p_{b1})}b_{1,0,0}, ~~~j \in (M,L), ~~~k \in (1,W_{1,M}-1)\\
b_{IDLE1} = \frac{1-\rho_{1}}{p_{a1}}b_{1,0,0}\\ 
\end{cases}
\end{equation}

The probability that the higher priority AC0 successfully transmits a packet is equal to the stationary probability of being in state 0. Furthermore, for AC1, $\omega_1$ can be viewed as the sum of the stationary probabilities of all the possible backoff states. $\omega_0$ and $\omega_1$ are given by equation~\ref{10}.

\begin{equation}
\begin{cases}
    
\omega_0  = b_{0,0} = \left[ 1 + \frac{(W_{0,0} - 1)}{2 (1 - p_{b0})} + \frac{1 - \rho_0}{p_{a0}} \right ]^{-1} \\
    

\omega_1 = \sum_{j=0}^{L} b_{1,j,0} = \frac{1 - p_{v1}^{L + 1}}{1 - p_{v1}} {\left[ \frac{1 - p_{v1}^{L + 1}}{1 - p_{v1}} + \frac{W_{1,0} - 1}{2 (1 - p_{b1})} + \frac{1}{2(1-p_{b1})} \left( \frac{W_{1,0} (2p_{v1}) [1 - (2 p_{v1})^{M_1}]}{1 - 2 p_{v1}} - \frac{p_{v1} (1-p_{v1}^{M_1}) }{1 - p_{v1}} + \frac{(2^{M_1}W_{1,0}-1) (1 - p_{v1}^{L - M_1}) p_{v1}^{M_1 + 1} }{1 - p_{v1}} \right) + \frac{1 - \rho_1}{p_{a1}} \right]^{-1}} 
\end{cases} \label{10}
\end{equation}   

The channel access parameters are given in Table~\ref{tab:Table 3} and the other traffic and network parameters are presented in Table~\ref{tab:Table 4}. The parameters are similar to the ones used by authors in~\cite{highway_reliability}. The overall delay experienced by a packet would comprise of two components: (i) the queuing and channel access delay, (ii) the propagation delay. The former is the sojourn time spent by a packet in the MAC layer, waiting in the AC queue and attempting to access the channel through contention and backoff procedure. The latter is the time spent by the channel in the physical wireless channel. In our traffic setup, we consider a deterministic propagation delay of $2\mu$s, which forms a very small component of the overall delay observed. Consequently, one may conclude that the MAC access delay forms the larger component of the packet delay and its PGF is given by~\cite{OnMACAccess}:
\begin{equation}
    P_{Tsi}(z) = \sum \limits_{k = 0}^{\infty} q_{i,k}z^{t_{{s}_{i,k}}}
\end{equation}
Thus, the delay experienced by a packet can be termed as the MAC access delay and used as a performance metric for evaluating the IEEE 802.11p protocol.
The mean of the MAC access delay can be evaluated by solving equations~\eqref{9} and ~\eqref{10} and is given by ~\cite{highway_reliability}:
\begin{equation}
    T_{Si} = \frac{dP_{Tsi}(z)}{dz} \mid_{z = 1} = {P_{Tsi}}^{\prime} (1) ~~, i = 0, 1 \label{12}
\end{equation}
where, the Probability Generating Function (PGF) $P_{Tsi}(z)$ is given as~\cite{highway_reliability}:
\begin{equation}
\begin{cases}
    P_{Ts0}(z) = \frac{TR(z)}{W_{0,0}} \sum \limits_{k = 0}^{W_{0,0} - 1} [H_0(z)]^k \\
\begin{gathered}
    P_{Ts1}(z) = (1 - p_{v1}) TR(z) \sum \limits_{n = 0}^{L} \left[ p_{v1}^{n} \prod \limits_{j = 0}^{n} B_{1,j}(z) \right] \\ + p_{v1}^{L + 1} \prod \limits_{j = 0}^{L} B_{1,j}(z)
\end{gathered}
\end{cases} \label{13}
\end{equation}

PGF $TR(z)$, $H_i(z)$ and $B_{1,j}(z)$ are given by~\cite{highway_reliability}:
\begin{equation}
    TR(z) = z^{T_{tr}} \label{14}
\end{equation}
\begin{equation}
    H_i(z) = (1- p_{bi}) z^{\sigma} + p_{bi} z^{T_{tr} + AIFS[i]} ~~, i = 0, 1 \label{15}
\end{equation}
\begin{equation}
   B_{1,j}(z) = \begin{cases}
    \frac{1}{W_{1,j}} \sum \limits_{k = 0}^{W_{1,j} - 1} [H_1(z)]^k ~~, j \in (1, M_1) \\
    \frac{1}{W_{1,M_1}} \sum \limits_{k = 0}^{W_{1,M_1} - 1} [H_1(z)]^k ~~, j \in (M_1, L)
\end{cases} \label{16}
\end{equation}

The standard deviation of the MAC access delay can be obtained as~\cite{highway_reliability}:
\begin{equation}
    D_{Tsi} = {P_{Tsi}}^{\prime \prime} (1) + {P_{Tsi}}^{\prime} (1) - \left[{P_{Tsi}}^{\prime} (1)\right]^2~~, i = 0, 1 \label{17}
\end{equation}
Figure~\ref{meandelay} shows the mean MAC access delay obtained for various values of headway. Simulation of the protocol is carried on the MATLAB 2021a. For the simulation scenario, we consider vehicles distributed evenly on a road and spaced by a desired headway (as in Figure~\ref{fig:my_label}). The parameters of 802.11p are set up in accordance with the IEEE 802.11p standard~\cite{standard}. From Figure~\ref{meandelay}, it is clear that the MAC access delay obtained using the simulation matches with the analytical results. As expected, the mean delay for the AC0 traffic is lower than that of the AC1 traffic. It can be observed that the mean access delay decreases as a function of headway. At a given instant, only a single vehicle can access the communication channel to broadcast packets to the vehicles which are in its transmission range. The access delay is composed of the backoff time (which is a function of $p_{bi}$) and the transmission time of which the transmission time is deterministic. When the headway is less, the vehicles are close to each other. With this $\rho_i$ increases. This technically means that the mac queue has more packets to serve. This increases the backoff time and the contention for the channel hence, increasing the mac access delay.\\
Figure~\ref{stddeviation} illustrates the standard deviation of the MAC access delay. The simulation results match with the analytical results. From the figure, it is evident that the deviation is more for lower values of headway. It is therefore important to study the distribution of the MAC access delay as the mean value may not give us the true picture about the actual traffic condition.

By utilizing the characteristics of PGF, one can obtain the Probability Mass Function~(PMF) of the MAC access delay as:
\begin{equation}
    p_k(t_{si}) = P\{t_{si}=k\} = \frac{{P_{T_{si}}^{(k)}} {(0)}}{k!}
\end{equation}

\begin{figure*}[!h]
\centering

\begin{subfigure}{0.325\textwidth}
\centering
\begin{tikzpicture}
 
\begin{axis}[
    width = \textwidth,
    height = \textwidth,
    legend pos = south east,
    xlabel = {Access Delay (ms)},
    ylabel = {Probability},
    xmin = 0, xmax = 60,
    ymin = 0.0, ymax = 1.0]
    \addplot[
        domain = 1.421:60,
        samples = 200,
        smooth,
        very thick,
        red,
    ] {1-exp(-0.2918*(x-1.421))};
\addlegendentry{\(AC0\)}
\addplot[
        domain = 1.421:60,
        samples = 200,
        smooth,
        very thick,
        blue,
    ] {1-exp(-0.1024*(x-1.421))};
\addlegendentry{\(AC1\)}
\end{axis}
\end{tikzpicture}
\caption{Headway = 3m}
\end{subfigure}
\hfill
\begin{subfigure}{0.325\textwidth}
\centering
\begin{tikzpicture}
 
\begin{axis}[
    width = \textwidth,
    height = \textwidth,
    legend pos = south east,
    xlabel = {Access Delay (ms)},
    ylabel = {Probability},
    xmin = 0, xmax = 60,
    ymin = 0.0, ymax = 1.0]
    \addplot[
        domain = 1.421:60,
        samples = 200,
        smooth,
        very thick,
        red,
    ] {1-exp(-0.4224*(x-1.421))};
\addlegendentry{\(AC0\)}
\addplot[
        domain = 1.421:60,
        samples = 200,
        smooth,
        very thick,
        blue,
    ] {1-exp(-0.1351*(x-1.421))};
\addlegendentry{\(AC1\)}
\end{axis}
\end{tikzpicture}
\caption{Headway = 5m}
\end{subfigure}
\hfill
\begin{subfigure}{0.325\textwidth}
\centering
\begin{tikzpicture}
 
\begin{axis}[
    width = \textwidth,
    height = \textwidth,
    legend pos = south east,
    xlabel = {Access Delay (ms)},
    ylabel = {Probability},
    xmin = 0, xmax = 60,
    ymin = 0.0, ymax = 1.0]
    \addplot[
        domain = 1.421:60,
        samples = 200,
        smooth,
         very thick,
        red,
    ] {1-exp(-0.674*(x-1.421))};
\addlegendentry{\(AC0\)}
\addplot[
        domain = 1.421:60,
        samples = 200,
        smooth,
        very thick,
        blue,
    ] {1-exp(-0.2109*(x-1.421))};
\addlegendentry{\(AC1\)}
\end{axis}
\end{tikzpicture}
\caption{Headway = 10m}
\end{subfigure}
\caption{CDF observed when AC0 packets are generated as per Linear model.}
\label{CDF-linear}
\end{figure*}
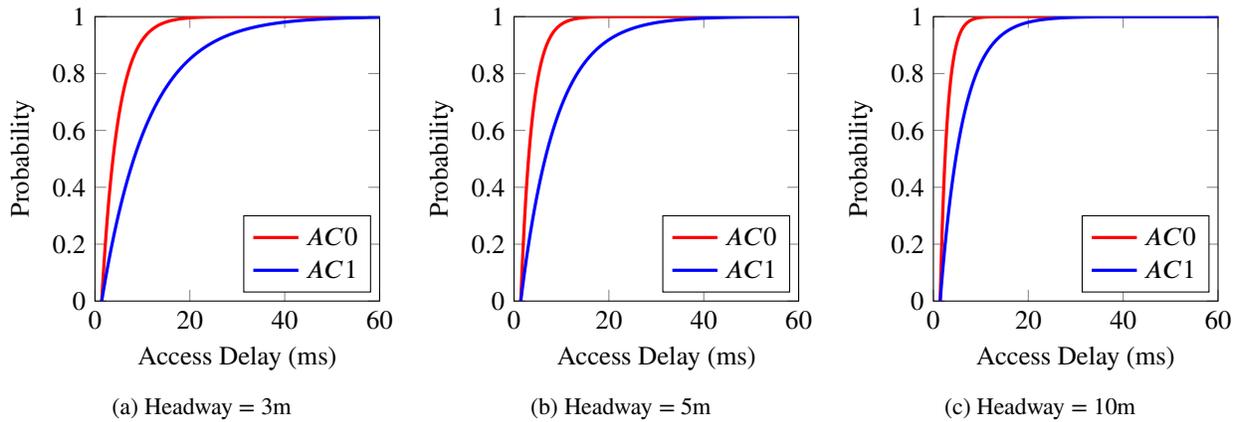

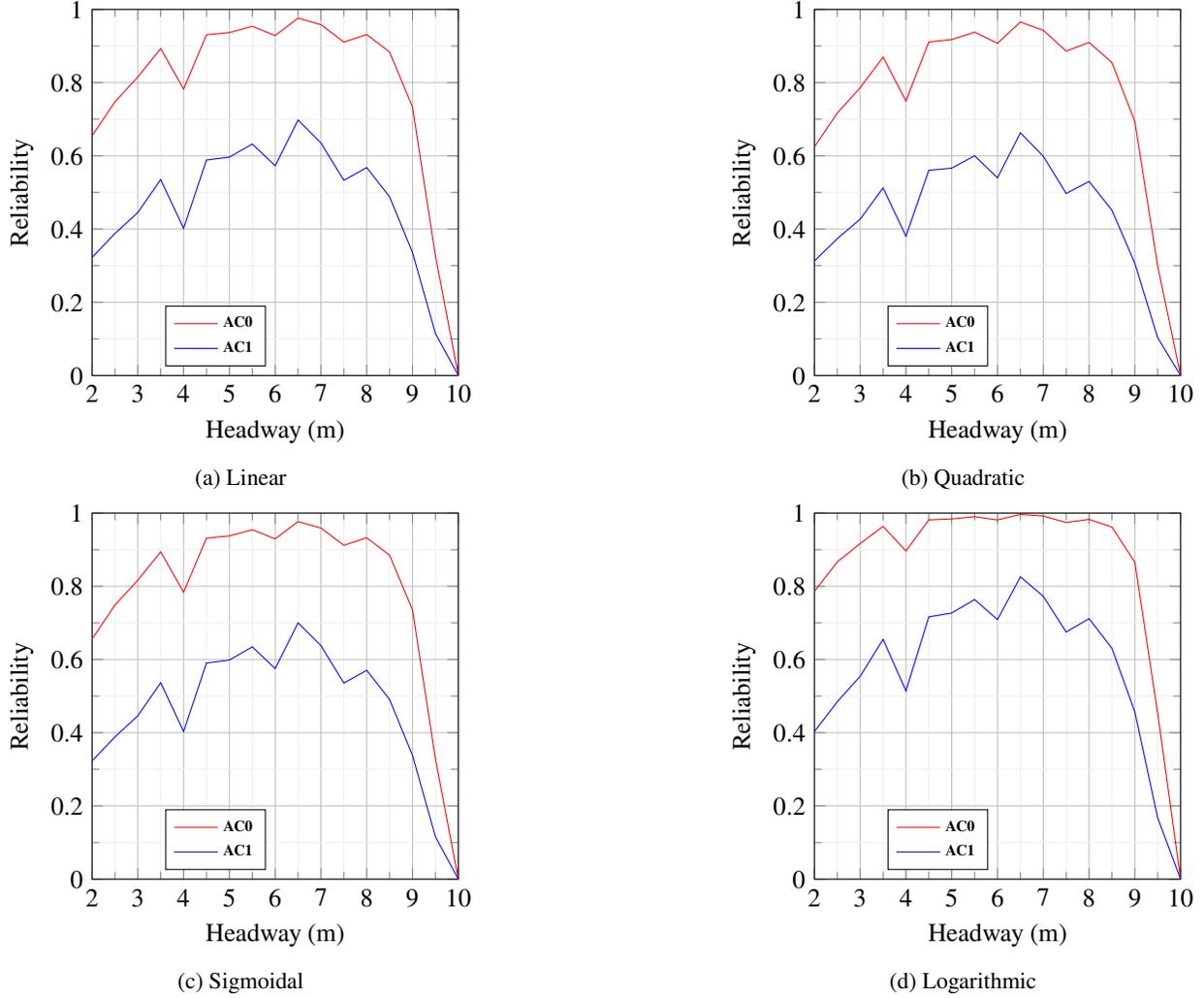
\begin{figure*}[!h]
\centering
\begin{subfigure}{0.40\textwidth}
\centering
\pgfplotstableread{rel_f1.dat}{\table}
\begin{tikzpicture}
\begin{axis}[
    xlabel = {Headway (m)},
    ylabel = {Reliability},
    xmin = 2, xmax = 10,
    ymin = 0, ymax = 1,
    xtick distance = 1,
    ytick distance = 0.2,
    grid = both,
    minor tick num = 1,
    major grid style = {lightgray},
    minor grid style = {lightgray!25},
    width = \textwidth,
    height = \textwidth,
    legend cell align = {right},
    legend style={at={(0.2,0.025)}, anchor=south west,font=\tiny}
]
 
\addplot[red] table [x = {x}, y = {y1}] {\table};
 \addplot[blue] table [x = {x}, y = {y2}] {\table};

 \legend{
     \textbf{AC0},
     \textbf{AC1},
    }
 
\end{axis}
 \end{tikzpicture}
\caption{Linear}
\end{subfigure}
\hfill
\begin{subfigure}{0.40\textwidth}
\centering
\pgfplotstableread{rel_f2.dat}{\table}
\begin{tikzpicture}
\begin{axis}[
    xlabel = {Headway (m)},
    ylabel = {Reliability},
    xmin = 2, xmax = 10,
    ymin = 0, ymax = 1,
    xtick distance = 1,
    ytick distance = 0.2,
    grid = both,
    minor tick num = 1,
    major grid style = {lightgray},
    minor grid style = {lightgray!25},
    width = \textwidth,
    height = \textwidth,
    legend cell align = {right},
    legend style={at={(0.2,0.025)}, anchor=south west,font=\tiny}
]
 
\addplot[red] table [x = {x}, y = {y1}] {\table};
 \addplot[blue] table [x = {x}, y = {y2}] {\table};

 \legend{
     \textbf{AC0}, 
      \textbf{AC1},
    }
 
\end{axis}
 \end{tikzpicture}
\caption{Quadratic}
\end{subfigure}
\hfill
\begin{subfigure}{0.40\textwidth}
\centering
\pgfplotstableread{rel_f3.dat}{\table}
\begin{tikzpicture}
\begin{axis}[
    xlabel = {Headway (m)},
    ylabel = {Reliability},
    xmin = 2, xmax = 10,
    ymin = 0, ymax = 1,
    xtick distance = 1,
    ytick distance = 0.2,
    grid = both,
    minor tick num = 1,
    major grid style = {lightgray},
    minor grid style = {lightgray!25},
    width = \textwidth,
    height = \textwidth,
    legend cell align = {right},
    legend style={at={(0.2,0.025)}, anchor=south west,font=\tiny}
]
 
\addplot[red] table [x = {x}, y = {y1}] {\table};
 \addplot[blue] table [x = {x}, y = {y2}] {\table};

 \legend{
     \textbf{AC0}, 
     \textbf{AC1},
    }
 
\end{axis}
 \end{tikzpicture}
\caption{Sigmoidal}
\end{subfigure}
\hfill
\begin{subfigure}{0.40\textwidth}
\centering
\pgfplotstableread{rel_f4.dat}{\table}
\begin{tikzpicture}
\begin{axis}[
    xlabel = {Headway (m)},
    ylabel = {Reliability},
    xmin = 2, xmax = 10,
    ymin = 0, ymax = 1,
    xtick distance = 1,
    ytick distance = 0.2,
    grid = both,
    minor tick num = 1,
    major grid style = {lightgray},
    minor grid style = {lightgray!25},
    width = \textwidth,
    height = \textwidth,
    legend cell align = {right},
    legend style={at={(0.2,0.025)}, anchor=south west,font=\tiny}
]
 
\addplot[red] table [x = {x}, y = {y1}] {\table};
 \addplot[blue] table [x = {x}, y = {y2}] {\table};

 \legend{
     \textbf{AC0},
     \textbf{AC1},
    }
 
\end{axis}
 \end{tikzpicture}
\caption{Logarithmic}
\end{subfigure}

\caption{Reliability observed for various models of AC0 traffic when vehicular mobility follows FVD model.}
\label{reliability}
\end{figure*}

\begin{figure*}[!h]
\centering
\begin{subfigure}{0.40\textwidth}
\centering
\pgfplotstableread{ovm_rel_f1.dat}{\table}
\begin{tikzpicture}
\begin{axis}[
    xlabel = {Headway (m)},
    ylabel = {Reliability},
    xmin = 2, xmax = 10,
    ymin = 0, ymax = 1,
    xtick distance = 1,
    ytick distance = 0.2,
    grid = both,
    minor tick num = 1,
    major grid style = {lightgray},
    minor grid style = {lightgray!25},
    width = \textwidth,
    height = \textwidth,
    legend cell align = {right},
    legend style={legend pos=south east,font=\tiny}
]
 
\addplot[red] table [x = {x}, y = {y1}] {\table};
 \addplot[blue] table [x = {x}, y = {y2}] {\table};

 \legend{
     \textbf{AC0},
     \textbf{AC1},
    }
 
\end{axis}
 \end{tikzpicture}
\caption{Linear}
\end{subfigure}
\hfill
\begin{subfigure}{0.40\textwidth}
\centering
\pgfplotstableread{ovm_rel_f2.dat}{\table}
\begin{tikzpicture}
\begin{axis}[
    xlabel = {Headway (m)},
    ylabel = {Reliability},
    xmin = 2, xmax = 10,
    ymin = 0, ymax = 1,
    xtick distance = 1,
    ytick distance = 0.2,
    grid = both,
    minor tick num = 1,
    major grid style = {lightgray},
    minor grid style = {lightgray!25},
    width = \textwidth,
    height = \textwidth,
    legend cell align = {right},
    legend style={legend pos=south east,font=\tiny}
]
 
\addplot[red] table [x = {x}, y = {y1}] {\table};
 \addplot[blue] table [x = {x}, y = {y2}] {\table};

 \legend{
     \textbf{AC0}, 
      \textbf{AC1},
    }
 
\end{axis}
 \end{tikzpicture}
\caption{Quadratic}
\end{subfigure}
\hfill
\begin{subfigure}{0.40\textwidth}
\centering
\pgfplotstableread{ovm_rel_f3.dat}{\table}
\begin{tikzpicture}
\begin{axis}[
    xlabel = {Headway (m)},
    ylabel = {Reliability},
    xmin = 2, xmax = 10,
    ymin = 0, ymax = 1,
    xtick distance = 1,
    ytick distance = 0.2,
    grid = both,
    minor tick num = 1,
    major grid style = {lightgray},
    minor grid style = {lightgray!25},
    width = \textwidth,
    height = \textwidth,
    legend cell align = {right},
    legend style={legend pos=south east,font=\tiny}
]
 
\addplot[red] table [x = {x}, y = {y1}] {\table};
 \addplot[blue] table [x = {x}, y = {y2}] {\table};

 \legend{
     \textbf{AC0}, 
     \textbf{AC1},
    }
 
\end{axis}
 \end{tikzpicture}
\caption{Sigmoidal}
\end{subfigure}
\hfill
\begin{subfigure}{0.40\textwidth}
\centering
\pgfplotstableread{ovm_rel_f4.dat}{\table}
\begin{tikzpicture}
\begin{axis}[
    xlabel = {Headway (m)},
    ylabel = {Reliability},
    xmin = 2, xmax = 10,
    ymin = 0, ymax = 1,
    xtick distance = 1,
    ytick distance = 0.2,
    grid = both,
    minor tick num = 1,
    major grid style = {lightgray},
    minor grid style = {lightgray!25},
    width = \textwidth,
    height = \textwidth,
    legend cell align = {right},
    legend style={legend pos=south east,font=\tiny}
]
 
\addplot[red] table [x = {x}, y = {y1}] {\table};
 \addplot[blue] table [x = {x}, y = {y2}] {\table};

 \legend{
     \textbf{AC0},
     \textbf{AC1},
    }
 
\end{axis}
 \end{tikzpicture}
\caption{Logarithmic}
\end{subfigure}

\caption{Reliability observed for various models of AC0 traffic when vehicular mobility follows MOVM model.}
\label{reliability-ovm}
\end{figure*}
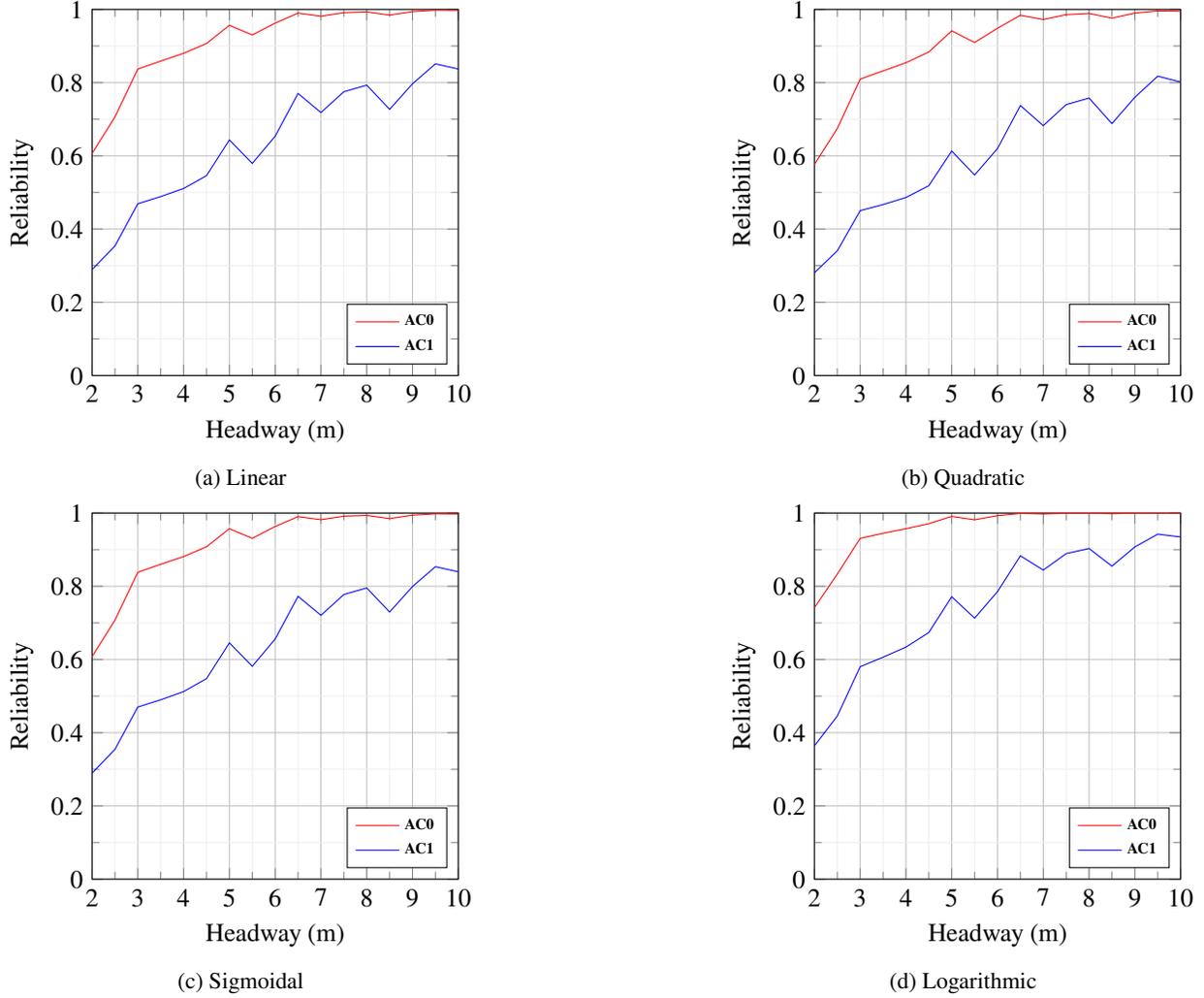

Once we obtain the PMF for each headway value, it is easy to obtain the Cumulative Distribution Function (CDF). For this, we use the curve fitting approach wherein we fit the curve according to shifted exponential distribution \cite{OnMACAccess}. Figure~\ref{CDF-linear} illustrates the CDF for Linear model of AC0 generation rate. The CDF plots for the other three models follow similar trend as the Linear model. The equations for CDF can be obtained in terms of headway. Table~\ref{CDF-headway} illustrates the CDF for different models of AC0 generation rate in terms of headway. For ensuring the stability of the platoon, it is desirable that every packet be delivered within the critical time delay. Therefore, one may quantify the performance of the IEEE 802.11p protocol in terms of its ability to ensure packet delivery times lesser than the critical delay. From the CDF, we can obtain the reliability of the protocol which is defined as the probability that a packet is delivered within a duration defined by the critical delay. The reliability obtained for various values of headway is shown in Figure~\ref{reliability} and Figure~\ref{reliability-ovm} when vehicles in the platoon follow FVD and MOVM  model respectively.


\begin{table}[!h]
\centering
\caption{EDCA access parameters}
\label{tab:Table 3}
\begin{tabular}{ccccc}
\toprule
AC & CWmin & CWmax & AIFSN \\
\toprule
\toprule
$1$ & 15 & 31 & $3$ \\
\bottomrule
\end{tabular}
\end{table}

\begin{table}[!h]
\centering
\caption{Traffic and network parameter settings}
\label{tab:Table 4}
\begin{tabular}{cccc}
\toprule
Parameter & Value & Parameter & Value \\
\toprule
\toprule
Basic rate & $1$ Mbps & Data rate & $3$ Mbps \\
PHY header & $48$ bits & MAC header & $112$ bits \\
Transmission range & $500$ m & Basic rate & $1$ Mbps \\
Carrier sensing range & $700$ m & Slot time & $13$ $\mu$s \\
$E[P]$ & $500$ bytes & SIFS & $32$ $\mu$s \\
Retry limit & $2$ & Propagation delay & $2$ $\mu$s \\
\bottomrule
\end{tabular}
\end{table}


\begin{figure*}[!h]
\centering
\begin{subfigure}{0.40\textwidth}
\centering
\pgfplotstableread{delay_f1.dat}{\table}
\begin{tikzpicture}
\begin{axis}[
    xlabel = {Headway (m)},
    ylabel = {Mean MAC Access Delay (ms)},
    xmin = 2, xmax = 10,
    ymin = 0, ymax = 14,
    xtick distance = 1,
    ytick distance = 2,
    grid = both,
    minor tick num = 1,
    major grid style = {lightgray},
    minor grid style = {lightgray!25},
    width = \textwidth,
    height = \textwidth,
    legend cell align = {right},
    legend style={legend pos=north east,font=\tiny}
]
 
\addplot[red] table [x = {x}, y = {y1}] {\table};
 \addplot[blue] table [x = {x}, y = {y2}] {\table};
\addplot[red, only marks] table [x = {x}, y = {y3}] {\table};
 \addplot[blue, only marks, mark = square,] table [x ={x}, y = {y4}] {\table};
 \legend{
    \textbf{AC0 Analysis}, 
     \textbf{ AC1 Analysis},
    \textbf{ AC0 Simulation},
    \textbf{ AC1 Simulation},
    }
 
\end{axis}
 \end{tikzpicture}
\caption{Linear}
\end{subfigure}
\hfill
\begin{subfigure}{0.40\textwidth}
\centering
\pgfplotstableread{delay_f2.dat}{\table}
\begin{tikzpicture}
\begin{axis}[
    xlabel = {Headway (m)},
    ylabel = {Mean MAC Access Delay (ms)},
    xmin = 2, xmax = 10,
    ymin = 0, ymax = 14,
    xtick distance = 1,
    ytick distance = 2,
    grid = both,
    minor tick num = 1,
    major grid style = {lightgray},
    minor grid style = {lightgray!25},
    width = \textwidth,
    height = \textwidth,
    legend cell align = {right},
    legend style={legend pos=north east,font=\tiny}
]
 
\addplot[red] table [x = {x}, y = {y1}] {\table};
 \addplot[blue] table [x = {x}, y = {y2}] {\table};
\addplot[red, only marks] table [x = {x}, y = {y3}] {\table};
 \addplot[blue, only marks, mark = square,] table [x ={x}, y = {y4}] {\table};
 \legend{
     \textbf{AC0 Analysis}, 
     \textbf{ AC1 Analysis},
    \textbf{ AC0 Simulation},
    \textbf{ AC1 Simulation},
    }
 
\end{axis}
 \end{tikzpicture}
\caption{Quadratic}
\end{subfigure}
\hfill
\begin{subfigure}{0.40\textwidth}
\centering
\pgfplotstableread{delay_f3.dat}{\table}
\begin{tikzpicture}
\begin{axis}[
    xlabel = {Headway (m)},
    ylabel = {Mean MAC Access Delay (ms)},
    xmin = 2, xmax = 10,
    ymin = 0, ymax = 14,
    xtick distance = 1,
    ytick distance = 2,
    grid = both,
    minor tick num = 1,
    major grid style = {lightgray},
    minor grid style = {lightgray!25},
    width = \textwidth,
    height = \textwidth,
    legend cell align = {right},
    legend style={legend pos=north east,font=\tiny}
]
 
\addplot[red] table [x = {x}, y = {y1}] {\table};
 \addplot[blue] table [x = {x}, y = {y2}] {\table};
\addplot[red, only marks] table [x = {x}, y = {y3}] {\table};
 \addplot[blue, only marks, mark = square,] table [x ={x}, y = {y4}] {\table};
 \legend{
     \textbf{AC0 Analysis}, 
     \textbf{ AC1 Analysis},
    \textbf{ AC0 Simulation},
    \textbf{ AC1 Simulation},
    }
 
\end{axis}
 \end{tikzpicture}
\caption{Sigmoidal}
\end{subfigure}
\hfill
\begin{subfigure}{0.40\textwidth}
\centering
\pgfplotstableread{delay_f4.dat}{\table}
\begin{tikzpicture}
\begin{axis}[
    xlabel = {Headway (m)},
    ylabel = {Mean MAC Access Delay (ms)},
    xmin = 2, xmax = 10,
    ymin = 0, ymax = 14,
    xtick distance = 1,
    ytick distance = 2,
    grid = both,
    minor tick num = 1,
    major grid style = {lightgray},
    minor grid style = {lightgray!25},
    width = \textwidth,
    height = \textwidth,
    legend cell align = {right},
    legend style={legend pos=north east,font=\tiny}
]
 
\addplot[red] table [x = {x}, y = {y1}] {\table};
 \addplot[blue] table [x = {x}, y = {y2}] {\table};
\addplot[red, only marks] table [x = {x}, y = {y3}] {\table};
 \addplot[blue, only marks, mark = square,] table [x ={x}, y = {y4}] {\table};
 \legend{
     \textbf{AC0 Analysis}, 
     \textbf{ AC1 Analysis},
    \textbf{ AC0 Simulation},
    \textbf{ AC1 Simulation},
    }
 
\end{axis}
 \end{tikzpicture}
\caption{Logarithmic}
\end{subfigure}

\caption{Mean MAC Access Delay for various models of packet generation rate for AC0 traffic based on the gap-acceptance probability of motorised two wheelers for each headway value.}
\label{meandelay}
\end{figure*}
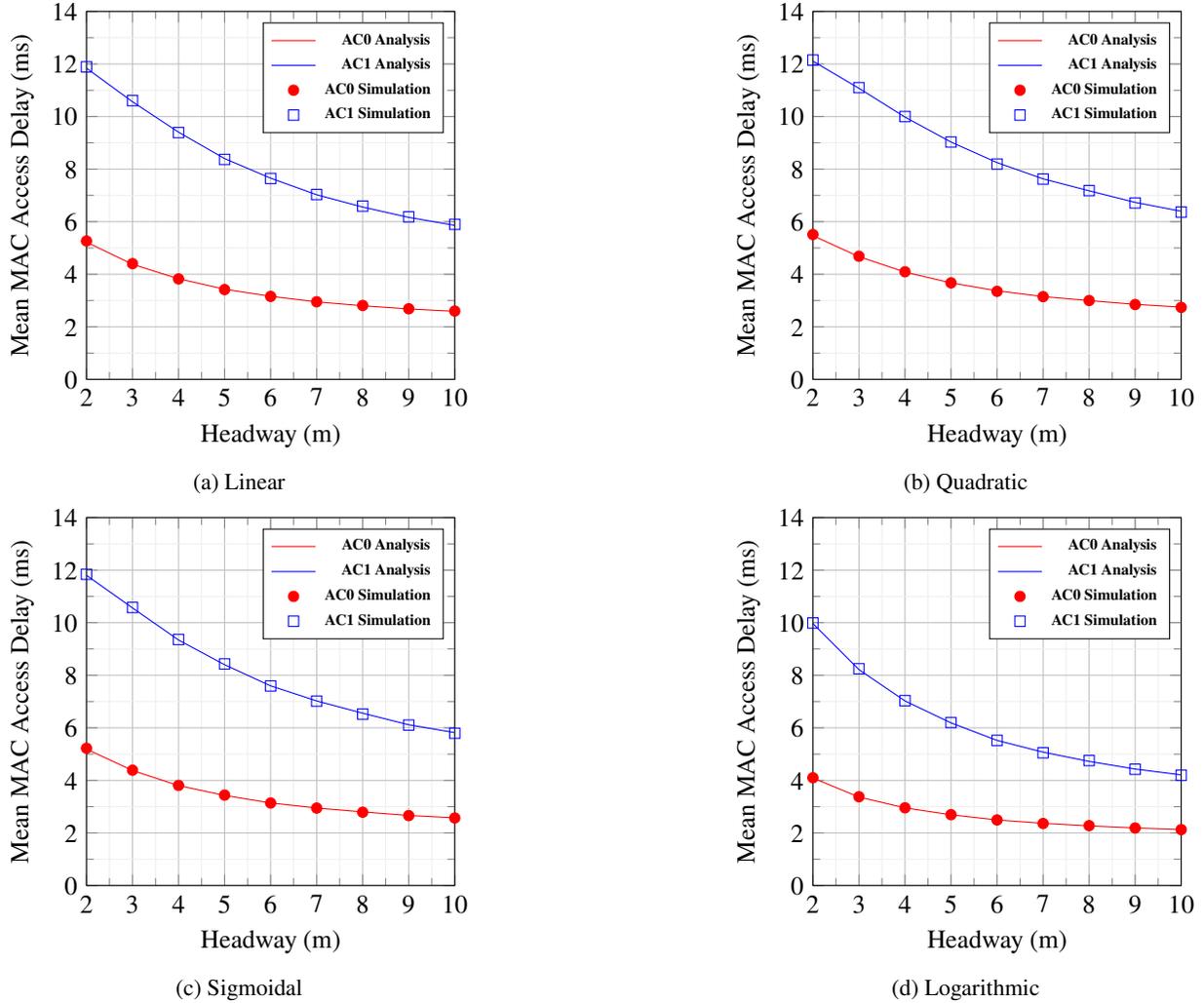

\begin{figure*}[!h]
\centering
\begin{subfigure}{0.40\textwidth}
\centering
\pgfplotstableread{std_f1.dat}{\table}
\begin{tikzpicture}
\begin{axis}[
    xlabel = {Headway (m)},
    ylabel = {Standard Deviation (ms)},
    xmin = 2, xmax = 10,
    ymin = 0, ymax = 7,
    xtick distance = 1,
    ytick distance = 1,
    grid = both,
    minor tick num = 1,
    major grid style = {lightgray},
    minor grid style = {lightgray!25},
    width = \textwidth,
    height = \textwidth,
    legend cell align = {right},
    legend style={legend pos=north east,font=\tiny}
]
 
\addplot[red] table [x = {x}, y = {y1}] {\table};
 \addplot[blue] table [x = {x}, y = {y2}] {\table};
\addplot[red, only marks] table [x = {x}, y = {y3}] {\table};
 \addplot[blue, only marks, mark = square,] table [x ={x}, y = {y4}] {\table};
 \legend{
     \textbf{AC0 Analysis},
     \textbf{ AC1 Analysis},
    \textbf{ AC0 Simulation},
    \textbf{ AC1 Simulation},
    }
 
\end{axis}
 \end{tikzpicture}
\caption{Linear}
\end{subfigure}
\hfill
\begin{subfigure}{0.40\textwidth}
\centering
\pgfplotstableread{std_f2.dat}{\table}
\begin{tikzpicture}
\begin{axis}[
    xlabel = {Headway (m)},
    ylabel = {Standard Deviation (ms)},
    xmin = 2, xmax = 10,
    ymin = 0, ymax = 7,
    xtick distance = 1,
    ytick distance = 1,
    grid = both,
    minor tick num = 1,
    major grid style = {lightgray},
    minor grid style = {lightgray!25},
    width = \textwidth,
    height = \textwidth,
    legend cell align = {right},
    legend style={at={(0.99,0.99)}, anchor=north east,font=\tiny}
]
 
\addplot[red] table [x = {x}, y = {y1}] {\table};
 \addplot[blue] table [x = {x}, y = {y2}] {\table};
\addplot[red, only marks] table [x = {x}, y = {y3}] {\table};
 \addplot[blue, only marks, mark = square,] table [x ={x}, y = {y4}] {\table};
 \legend{
     \textbf{AC0 Analysis}, 
      \textbf{ AC1 Analysis},
    \textbf{ AC0 Simulation},
    \textbf{ AC1 Simulation},
    }
 
\end{axis}
 \end{tikzpicture}
\caption{Quadratic}
\end{subfigure}
\hfill
\begin{subfigure}{0.40\textwidth}
\centering
\pgfplotstableread{std_f3.dat}{\table}
\begin{tikzpicture}
\begin{axis}[
    xlabel = {Headway (m)},
    ylabel = {Standard Deviation (ms)},
    xmin = 2, xmax = 10,
    ymin = 0, ymax = 7,
    xtick distance = 1,
    ytick distance = 1,
    grid = both,
    minor tick num = 1,
    major grid style = {lightgray},
    minor grid style = {lightgray!25},
    width = \textwidth,
    height = \textwidth,
    legend cell align = {right},
    legend style={legend pos=north east,font=\tiny}
]
 
\addplot[red] table [x = {x}, y = {y1}] {\table};
 \addplot[blue] table [x = {x}, y = {y2}] {\table};
\addplot[red, only marks] table [x = {x}, y = {y3}] {\table};
 \addplot[blue, only marks, mark = square,] table [x ={x}, y = {y4}] {\table};
 \legend{
     \textbf{AC0 Analysis}, 
     \textbf{ AC1 Analysis},
    \textbf{ AC0 Simulation},
    \textbf{ AC1 Simulation},
    }
 
\end{axis}
 \end{tikzpicture}
\caption{Sigmoidal}
\end{subfigure}
\hfill
\begin{subfigure}{0.40\textwidth}
\centering
\pgfplotstableread{std_f4.dat}{\table}
\begin{tikzpicture}
\begin{axis}[
    xlabel = {Headway (m)},
    ylabel = {Standard Deviation (ms)},
    xmin = 2, xmax = 10,
    ymin = 0, ymax = 7,
    xtick distance = 1,
    ytick distance = 1,
    grid = both,
    minor tick num = 1,
    major grid style = {lightgray},
    minor grid style = {lightgray!25},
    width = \textwidth,
    height = \textwidth,
    legend cell align = {right},
    legend style={legend pos=north east,font=\tiny}
]
 
\addplot[red] table [x = {x}, y = {y1}] {\table};
 \addplot[blue] table [x = {x}, y = {y2}] {\table};
\addplot[red, only marks] table [x = {x}, y = {y3}] {\table};
 \addplot[blue, only marks, mark = square,] table [x ={x}, y = {y4}] {\table};
 \legend{
     \textbf{AC0 Analysis},
     \textbf{ AC1 Analysis},
    \textbf{ AC0 Simulation},
    \textbf{ AC1 Simulation},
    }
 
\end{axis}
 \end{tikzpicture}
\caption{Logarithmic}
\end{subfigure}

\caption{Standard Deviation (ms) observed for various models of AC0 traffic.}
\label{stddeviation}
\end{figure*}
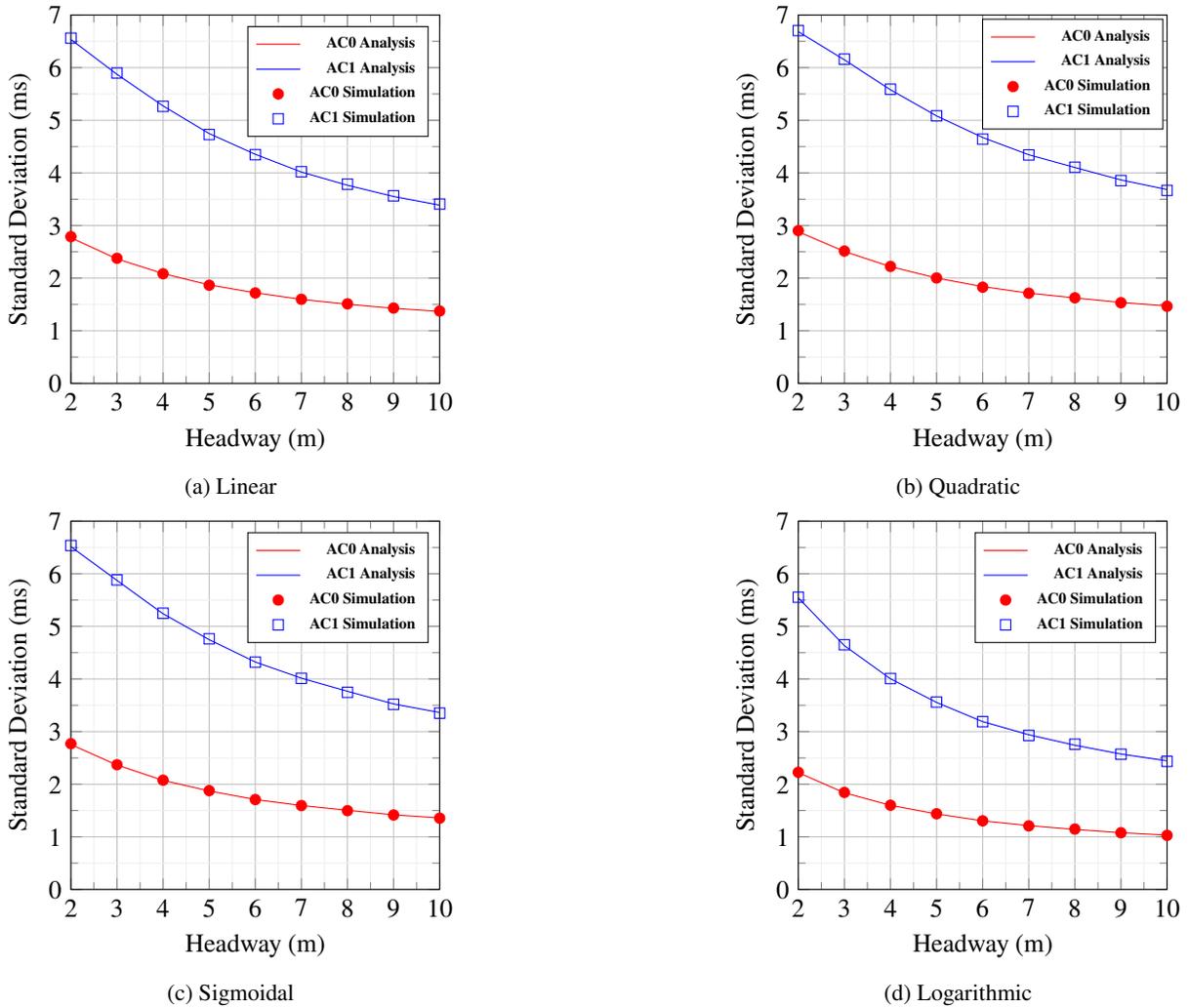

\begin{table*}[!h]
\centering
\caption{CDF obtained for different AC0 packet generating functions in terms of headway}
\begin{tabular}{cccc}
\toprule
& AC0 & AC1 \\
\toprule
\toprule
Linear & \(\displaystyle \begin{cases} 1 - \left( e^{-\left(0.0566 y^\ast + 0.1277 \right)\left(x - T_{tr}\right)} \right),~~x \geq T_{tr} \\ 0,~~x < T_{tr} \end{cases}\) & \(\displaystyle \begin{cases} 1 - \left( e^{-\left(0.0156 y^\ast + 0.057 \right)\left(x - T_{tr}\right)} \right),~~x \geq T_{tr} \\ 0,~~x < T_{tr} \end{cases}\) \\
\midrule
Quadratic & \(\displaystyle \begin{cases} 1 - \left( e^{-\left(0.0499 y^\ast + 0.1221 \right)\left(x - T_{tr}\right)} \right),~~x \geq T_{tr} \\ 0,~~x < T_{tr} \end{cases}\) & \(\displaystyle \begin{cases} 1 - \left( e^{-\left(0.0131 y^\ast + 0.0588 \right)\left(x - T_{tr}\right)} \right),~~x \geq T_{tr} \\ 0,~~x < T_{tr} \end{cases}\) \\
\midrule
Sigmoidal & \(\displaystyle \begin{cases} 1 - \left( e^{-\left(0.057 y^\ast + 0.1276 \right)\left(x - T_{tr}\right)} \right),~~x \geq T_{tr} \\ 0,~~x < T_{tr} \end{cases}\) & \(\displaystyle \begin{cases} 1 - \left( e^{-\left(0.0158 y^\ast + 0.0567 \right)\left(x - T_{tr}\right)} \right),~~x \geq T_{tr} \\ 0,~~x < T_{tr} \end{cases}\) \\
\midrule
Logarithmic & \(\displaystyle \begin{cases} 1 - \left( e^{-\left(0.0885 y^\ast + 0.1724 \right)\left(x - T_{tr}\right)} \right),~~x \geq T_{tr} \\ 0,~~x < T_{tr} \end{cases}\) & \(\displaystyle \begin{cases} 1 - \left( e^{-\left(0.0254 y^\ast + 0.0661 \right)\left(x - T_{tr}\right)} \right),~~x \geq T_{tr} \\ 0,~~x < T_{tr} \end{cases}\) \\
\bottomrule
\end{tabular}
\label{CDF-headway}
\end{table*}

\section{Conclusions and future work}
\label{sec:Contributions}

We studied the performance of IEEE 802.11p medium access protocol in the context of a heterogeneous traffic environment. We specifically investigated a traffic setup comprising a platoon of connected vehicles and human-driven motorised two-wheelers. FVD and MOVM were employed to model the mobility of the vehicles and the gap acceptance model was used to account for the likelihood of a two-wheeler cutting across the platoon. Using the above models, we configured the traffic setup and the rate of generation of data packets. Further, we conducted a local stability analysis of the FVD to determine a critical delay which serves as an upper bound on the delay experienced by a packet such that platoon stability can be maintained. We then obtained the mean MAC access delay and standard deviation for various values of the platoon headway.
CDF was obtained for different AC0 packet generating functions. Using the CDF, reliability of IEEE 802.11p protocol in context of heterogeneous traffic setup was derived.

The same study could be extended to incorporate vehicles of other types cutting across multiple platoons. In addition reliability can be computed by considering different mobility models (like the Intelligent driver model) of the vehicles in the platoon. While we carried out the performance analysis of IEEE  802.11p, other options like C-V2X, NR-V2X, or IEEE 802.11bd could be explored to analyze performance in heterogeneous traffic scenarios.

\bibliographystyle{IEEEtran}
\bibliography{References}

\begin{thebibliography}{10}
\providecommand{\url}[1]{#1}
\csname url@samestyle\endcsname
\providecommand{\newblock}{\relax}
\providecommand{\bibinfo}[2]{#2}
\providecommand{\BIBentrySTDinterwordspacing}{\spaceskip=0pt\relax}
\providecommand{\BIBentryALTinterwordstretchfactor}{4}
\providecommand{\BIBentryALTinterwordspacing}{\spaceskip=\fontdimen2\font plus
\BIBentryALTinterwordstretchfactor\fontdimen3\font minus
  \fontdimen4\font\relax}
\providecommand{\BIBforeignlanguage}[2]{{%
\expandafter\ifx\csname l@#1\endcsname\relax
\typeout{** WARNING: IEEEtran.bst: No hyphenation pattern has been}%
\typeout{** loaded for the language `#1'. Using the pattern for}%
\typeout{** the default language instead.}%
\else
\language=\csname l@#1\endcsname
\fi
#2}}
\providecommand{\BIBdecl}{\relax}
\BIBdecl

\bibitem{statistics2}
\BIBentryALTinterwordspacing
 [Online]. Available:
  \url{https://www.who.int/news-room/fact-sheets/detail/road-traffic-injuries}
\BIBentrySTDinterwordspacing

\bibitem{ieee1}
W.~Sun \emph{et~al.}, ``{Analytical study of the IEEE 802.11p EDCA
  mechanism},'' in \emph{2013 IEEE Intelligent Vehicles Symposium (IV)}.\hskip
  1em plus 0.5em minus 0.4em\relax IEEE, 2013, pp. 1428--1433.

\bibitem{ieee2}
S.~Cao and V.~C. Lee, ``{An accurate and complete performance modeling of the
  IEEE 802.11p MAC sublayer for VANET},'' \emph{Computer Communications}, vol.
  149, pp. 107--120, 2020.

\bibitem{highway}
Y.~Yao \emph{et~al.}, ``{Delay analysis and study of IEEE 802.11p based DSRC
  safety communication in a highway environment},'' in \emph{2013 Proceedings
  IEEE INFOCOM}, 2013, pp. 1591--1599.

\bibitem{highway_reliability}
Y.~{Y}ao \emph{et~al.}, ``Performance and {Reliability} {Analysis} of {IEEE}
  802.11p {Safety} {Communication} in a {Highway} {Environment},'' \emph{IEEE
  Transactions on Vehicular Technology}, vol.~62, no.~9, pp. 4198--4212, 2013.

\bibitem{OnMACAccess}
Y.~{Yao} \emph{et~al.}, ``On {MAC} {Access} {Delay} {Distribution} for {IEEE}
  802.11p {Broadcast} in {Vehicular} {Networks},'' \emph{IEEE Access}, vol.~7,
  pp. 149\,052--149\,067, 2019.

\bibitem{backofffreezing}
Q.~Wu \emph{et~al.}, ``{Performance analysis of IEEE 802.11p for continuous
  backoff freezing in IoV},'' \emph{Electronics}, vol.~8, no.~12, 2019.

\bibitem{hidden}
P.~Rathee \emph{et~al.}, ``{Performance Analysis of IEEE 802.11p in the
  Presence of Hidden Terminals},'' \emph{Wireless Personal Communications},
  vol.~89, pp. 61--78, 2016.

\bibitem{23}
H.~Peng \emph{et~al.}, ``Performance {Analysis} of {IEEE} 802.11p {DCF} for
  {Multiplatooning} {Communications} {With} {Autonomous} {Vehicles},''
  \emph{IEEE Transactions on Vehicular Technology}, vol.~66, no.~3, pp.
  2485--2498, 2017.

\bibitem{25}
K.~Xu \emph{et~al.}, ``Time-{Dependent} {Performance} {Analysis} of {IEEE}
  802.11p {Vehicular} {Networks},'' \emph{IEEE Transactions on Vehicular
  Technology}, vol.~65, no.~7, pp. 5637--5651, 2016.

\bibitem{1}
Q.~Wu \emph{et~al.}, ``Time-{Dependent} {Performance} {Analysis} of the
  802.11p-{Based} {Platooning} {Communications} {Under} {Disturbance},''
  \emph{IEEE Transactions on Vehicular Technology}, vol.~69, no.~12, pp.
  15\,760--15\,773, 2020.

\bibitem{9}
G.~K. Kamath \emph{et~al.}, ``The {Modified} {Optimal} {Velocity} {Model}:
  {Stability} {Analyses} and {Design} {Guidelines},'' \emph{IFAC Journal of
  Systems and Control}, vol.~2, pp. 18--32, 2017.

\bibitem{gap}
K.~Bhattacharyya \emph{et~al.}, ``Analytical and {Microsimulation} {Models} for
  {Investigating} {Vehicle} {Class}-{Specific} {Gap} {Acceptance} {Behavior} at
  {Urban} {Intersections} {With} {Nonlane}-{Based} {Traffic} {Operations},''
  \emph{Asian Transport Studies}, vol.~6, p. 100012, 2020.

\bibitem{16}
M.~Noor-A-Rahim \emph{et~al.}, ``Performance {Analysis} of {IEEE} 802.11p
  {Safety} {Message} {Broadcast} {With} and {Without} {Relaying} at {Road}
  {Intersection},'' \emph{IEEE Access}, vol.~6, pp. 23\,786--23\,799, 2018.

\bibitem{20}
J.~Zheng and Q.~Wu, ``Performance {Modeling} and {Analysis} of the {IEEE}
  802.11p {EDCA} {Mechanism} for {VANET},'' \emph{IEEE Transactions on
  Vehicular Technology}, vol.~65, no.~4, pp. 2673--2687, 2016.

\bibitem{27}
Z.~Tong \emph{et~al.}, ``A {Stochastic} {Geometry} {Approach} to the {Modeling}
  of {IEEE} 802.11 p for {Vehicular} {Ad} hoc {Networks},'' \emph{Submitted to
  IEEE Transactions on Vehicular Technology}, 2015.

\bibitem{6}
M.~Bando \emph{et~al.}, ``Dynamical {Model} of {Traffic} {Congestion} and
  {Numerical} {Simulation},'' \emph{Physical review E}, vol.~51, no.~2, p.
  1035, 1995.

\bibitem{8}
M.~{B}ando \emph{et~al.}, ``Analysis of {Optimal} {Velocity} {Model} {With}
  {Explicit} {Delay},'' \emph{Physical Review E}, vol.~58, no.~5, p. 5429,
  1998.

\bibitem{beregi2021connectivity}
S.~Beregi \emph{et~al.}, ``{Connectivity-based delay-tolerant control of
  automated vehicles: theory and experiments},'' \emph{IEEE Transactions on
  Intelligent Vehicles}, 2021.

\bibitem{standard}
``{IEEE Standard for Information technology-- Local and metropolitan area
  networks-- Specific requirements-- Part 11: Wireless LAN Medium Access
  Control (MAC) and Physical Layer (PHY) Specifications Amendment 6: Wireless
  Access in Vehicular Environments},'' pp. 1--51, 2010.

\end{thebibliography}

\end{document}